\documentstyle[12pt]{article}
\def\l{\lambda}
\def\m{\mu}
\def\n{\nu}

\def\o{\omega}
\def\d{\delta}
\def\e{\epsilon}

\def\be{\begin{equation}}
\def\ee{\end{equation}}
\def\p{\partial}
\def\ber{\begin{eqnarray}}
\def\eer{\end{eqnarray}}
\begin{document}

\begin{center}
{\Large\bf  Gauge Theories on Sphere and Killing Vectors }
\vskip 1 true cm
{\bf Rabin 
Banerjee}\footnote{On leave from S.N.Bose Natl. Ctr. for Basic Sciences,
Calcutta, India; e-mail:rabin@newton.skku.
ac.kr;  rabin@bose.res.in}
\vskip .8 true cm
BK21 Physics Research Division and Institute
of Basic Science,\\
 SungKyunKwan 
University,\\ Suwon 440- 746, Republic of Korea\end{center}
\bigskip

\centerline{\large\bf Abstract}
\medskip

We provide a general method for studying
 manifestly $O(n+1)$ covariant formulation of $p$-form  gauge
theories by stereographically projecting these theories, 
defined in flat Euclidean space, onto the
surface of a hypersphere. The gauge fields in the two descriptions are mapped by conformal
Killing vectors while conformal Killing spinors are necessary for the matter fields, allowing
for a very transparent analysis and compact 
presentation of results. General expressions for these Killing
vectors and spinors are given. The familiar results for a vector gauge
theory are reproduced.

\newpage

\section{Introduction}
 
\bigskip

Stereographic projection has important applications in both physics and mathematics. In general
terms, streographic projection from an $n$-dimensional sphere $(S)$, embedded in $(n+1)$- dimensional
flat Euclidean space, onto a plane tangent to $S$ at $x$ 
is the map that projects each 
point $P$ on
$ S$, to the intersection $P'$ of the line $Px'$ with the plane, where $x'$ is the
point opposite to $x$ on $S$. 
In mathematics, it gives a picture of $S^3$ as $R^3 \cup \{\infty\}$, that makes $S^3$ intuitively
more understandable than directly visualising it as the unit hypersphere $(x_1^2 + x_2^2 +x_3^2 +
x_4^2 = 1)$. Likewise. several models of hypebolic space are derived from more familiar models
(e.g. Poincare ball) by streographic projection \cite{T}.

In the realm of physics, manifestly $O(n+1)$ covariant formulation of vector gauge theories 
is done by stereographically projecting the usual gauge theory, defined on the Euclidean 
plane, onto 
the hypersphere\cite{A, DS}. It provides a deep understanding of the topological properties of a 
gauge theory\cite{JR, BP} including, for instance, the connection between the axial anomaly and the
index theorem\cite{NS}.

Streographic projection of a vector gauge theory is usually presented as a set of formulas
relating the gauge as well as matter sectors in the usual flat space and hyperspherical
formulations\cite{e}. Once this dictionary is established, the rest are technical details.
The derivation and significance of these formulas are, however, not particularly illuminating.
Also, their connection with the precise map among the coordinates of the plane and the sphere, 
obtained from a geometrical construction, is lost. Moreover, explicit results are given 
only for vector gauge theories in two and four dimensions; so it is not clear how far such
results are general and whether it is possible to analyse $p$-form gauge theories or
include higher dimensions. Indeed since antisymmetric tensor fields do occur in various contexts,
it is desirable, if not essential, to have a manifestly covariant formulation of such
theories for the precise reasons that led to the original treatment of the vector gauge
theory; namely, to avoid the ambiguities inherent in a flat space treatment by working
on a compact manifold.

In this paper we discuss a method, relying on first principles,
 for obtaining the various formulas and expressions providing the transition from the flat
space to the hypersphere. It is 
based on the observation that a stereographic projection is a conformal transformation.
We show that quantities in the gauge sector (like gauge fields, field strengths etc.)
in the two descriptions are related by rules similar to usual tensor analysis, with the
conformal Killing vectors playing the role of the metric. The explicit structures of these vectors
is derived by solving the Cartan-Killing equation. Using this formalism, results for 
a vector gauge theory are economically reproduced, apart from the fact that
they become quite transparent. It is then extended to obtain new 
results for a $p$-form gauge theory. Although the two form case has been actually done
in details, the treatment for higher forms parallels this example and is 
an obvious generalisation.
The analysis is done for arbitrary dimensions and nonabelian gauge groups have been 
considered.

The inclusion of matter sector is also possible within this scheme. Just as gauge fields
are related by conformal Killing vectors, the matter fields in the flat and hyperspherical
surfaces are  connected by conformal Killing spinors. We derive an equation where these Killing 
vectors are expressed as bilinear combinations of Killing spinors. The solution to this
equation yields the structure of these spinors. In this manner conformal 
Killing spinors on $S^n$
are given by a compact formula.

The paper is organised as follows: section 2 analyses the connection between stereographic 
projection and conformal Killing vectors, including a derivation of the latter from the
Cartan-Killing equation; sections 3 and 4 treat the covariant formulation of a vector
and a two form gauge theory, respectively, pointing out the differences
between the two; also, a new gauge symmetry for a two form theory on a hypersphere is
noted that does not have any analogue in the flat space; section 5 discusses the group contraction
of $SO(3)$ to $E(2)$ by means of the Killing vectors found here; section 6 contains 
concluding remarks.

\section{ Stereographic Projection and Killing Vectors: A First Principle Analysis}

A manifestly $O(n+1)$-covariant formulation of a gauge theory is attained 
by stereographically projecting $n$-dimensional Euclidean 
space onto the surface of a unit hypersphere embedded in $(n+1)$-dimensional 
Euclidean space. Introducing the $(n+1)$-dimensional coordinates on the 
unit hypersphere by $r_a (a=1,2,......(n+1))$ with $r_a r_a =1$, 
and the $n$-dimensional coordinates on the hyperplane by $x_\m (\m=1,2......n)$,
then a mapping from the south pole yields the familiar relations{\footnote{Latin labels run from 1 to $(n+1)$, while Greek labels run from 1 to $n$}},
\be
r_\m =\frac{2x_\m}{1+x^2}
\label{s0}
\ee
\be 
r_{(n+1)} = \frac{1-x^2}{1+x^2} 
\label{s1}
\ee

The gauge potentials on the sphere (denoted by a caret) and the 
conventional ones are likewise related by \cite{A, DS, JR},
\be
\hat A_\m =\frac{1+x^2}{2} A_\m - x_\m x_\n A^\n 
\label{s2}
\ee
\be 
\hat A_{(n+1)} = - x_\m A^\m
\label{s3}
\ee

It is easy to see that the inverse map is provided by,
\be
x_\m =\frac{r_\m}{1+r_3}
\label{inverse}
\ee
and,
\be
\frac{1+x^2}{2} A_\m = \hat A_\m - x_\m \hat A_{(n+1)}
\label{s5}
\ee
While (\ref{s0}) and (\ref{s1}) are a consequence of a straightforward geometrical exercise, the mapping among the potentials seems somewhat obscure, relegating it to a matter of definition. As far as degrees of freedom are concerned, these are preserved. The $(n+1)$-dimensional $\hat A$ variables are mapped to the $n$-dimensional $A$ variables, subjected to the constraint,
\be
r_a \hat A_a = 0
\label{s6}
\ee
implying that the $\hat A$-fields live on the tangent space of the hypersphere.

  We now provide a systematic 
derivation of (3) and (\ref{s3}), based on the symmetries of the problem. 
There is a mapping among the symmetries of the plane and the sphere 
(e.g. translations on the plane correspond to rotations on the sphere) 
that is captured by the relevant Killing vectors. 
Moreover, stereographic projection is known to be a 
conformal transformation. Even geometrically it can be proved
that stereographic projection from a sphere to a plane
is identical to an inversion
(which is a discrete transformation of the conformal
group) in a sphere of twice the radius \cite{T}.
 Variables on the sphere and plane should thus be related by 
conformal Killing vectors.  We may write this relation as,
 \be
 \hat A_a = K_a^\m A_\m + r_a\phi
 \label{s7}
 \ee
 where the conformal Killing vectors satisfy the transversality condition,
 \be
 r_a K_a^\m = 0
 \label{s8}
 \ee
 and an additional scalar field $\phi$, which is just the normal component of $\hat A_a$, is introduced,
 \be
 \phi=r_a\hat A_a
 \label{s9}
 \ee 

The $(n+1)$ components of $\hat A$ are expressed in terms of the $n$ components of $A$ plus a scalar degree of freedom. To simplify the analysis the scalar field is put to zero. It is straightforward to resurrect it by using the above equations. With the scalar field gone, $\hat A$ is now given by,
\be
\hat A_a = K_a^\m A_\m
\label{k}
\ee
and satisfies the condition (\ref{s6}), enabling a suitable comparison.

The conformal Killing vectors $K_a^\m$ are now determined. These should satisfy the Cartan-Killing equation which, specialised to a flat $n$-dimensional manifold, is given by,
\be
\p^\n K_a^\m +\p^\m K_a^\n = \frac{2}{n} \p^\l K_a^\l  \d_{\m\n}
\label{s10}
\ee

The most general solution for this equation (except for $n=2$) is given by \cite{FMS},
\be
K_a^\m= t_a^\m +\e_a x^\m + \omega_{a}^{\m\n} x_\n +\l_a^\m x^2 -2\lambda_a^ \sigma x_\sigma x^\m
\label{s11}
\ee
where $\o^{\m\n}= - \o^{\n\m}$.
The various transformations of the conformal group are characterised by the parameters appearing in the above equation; translations by $t$, dilitations by $\e$, rotations by $\o$ and inversions (or the special conformal transformations) by $\l$. Imposing the condition (\ref{s8}) and equating coefficients of terms with distinct powers of $x$, we find,
\ber
t_{(n+1)}^\m &=& 0 \\
2x_\nu  t_\n^\m + \o^{\m\n}_{(n+1)} x_\n + \e_{(n+1)}x^\m &=&0 \\
-x^2 t_{(n+1)}^\mu +
  2ix_\n\e_\n x^\m +2x_\n\o_{\n\sigma}^\m x^\sigma +\l^\m_{(n+1)} x^2 - 2\l^\sigma_{(n+1)} x_\sigma x^\m &=& 0\\
2x_\n\l_\n^\m x^2 - 4\l_\n^\sigma x_\n x_\sigma x^\m - x^2\e_{(n+1)}x^\m-\o_{(n+1)\n}^\m x^\n x^2 &=& 0\\
\l_{(n+1)}^\m x^2 - 2\l_{(n+1)}^\sigma x_\sigma x^\m &=&0
\label{s12}
\eer

Contracting the above equations (except the first one) by $x_\m$ yields,
\ber
2x_\mu x_\n t_\n^\m + \e_{(n+1)}x^2 &=&0 \\
2\e_\m x_\m - \l_{(n+1)}^\m x_\m &=& 0 \\
2\l_\n^\sigma x_\sigma x^\n + x^2 \e_{(n+1)} &=& 0 \\
x_\m x^2 \l_{(n+1)}^\m &=& 0
\label{s12a}
\eer

Since $x_\m$ is completely arbitrary, we find that $\l_{(n+1)}^\m = \e^\m =0$. Using this in (16) 
it is seen that $\o_{\n\sigma}^\m =0$. Equations (19) and (21) determine the symmetric parts of $t$ and $\l$, respectively, as,
\be
(t_\n^\m)_{symm.} =(\l_\n^\m)_{symm.} =  -\frac{1}{2}\d_\n^\m\e_{(n+1)} 
\label{s17}
\ee

Their antisymmetric parts are determined in terms of the antisymmetric tensor $\o$, on using (15) and (17),
so that the complete solution has the form,
\ber
t_\n^\m &=&   -\frac{1}{2}( \d_\n^\m\e_{(n+1)} +\o^\m_{(n+1)\n})\\
\l_\n^\m &=&  -\frac{1}{2}( \d_\n^\m\e_{(n+1)} -\o^\m_{(n+1)\n})
\label{s23}
\eer

The Cartan-Killing equation (\ref{s10}) is linear in the Killing vectors, which can thus be appropriately scaled. This freedom is exploited in setting the parameter of the scale transformations $\epsilon_{(n+1)}=-1$.  
That leaves only $\o^\m_{(n+1)\n}$ undetermined. This is set to zero by requiring that the Killing vectors are symmetric under interchange of $\m$ and $\n$, so that the rotational symmetry of the problem is preserved. Thus all the parameters have been fixed. The explicit structures of the Killing vectors, isolating the $(n+1)-th$ component are now written,
\be
K_\n^\m = \frac{1+x^2}{2}\d_\n^\m - x_\m x_\n 
\label{s18a}
\ee
\be
K_{(n+1)}^\m = - x_\m
\label{s18b}
\ee

Susequently we shall show that the standard forms of the Killing vectors, either on the sphere or on the plane, are recovered from this result.

With the above solution for the Killing vectors, eq. (\ref{k}) reproduces the defining relations for the potentials (3) and (\ref{s3}), thereby completing a systematic derivation for them.

There are two useful relations satisfied by these Killing vectors, namely;
\be
K_a^\m K_a^\n = \Big (\frac{1+x^2}{2}\Big)^2 \delta^{\m\n}
\label{k1}
\ee
and,
\be
K_a^\m K_b^\m = \Big(\frac{1+x^2}{2}\Big)^2 (\delta_{ab} - r_a r_b)
\label{k2}
\ee
Note that the conformal factor that relates the volume element on the hypersphere with that in the $n$-dimensional flat manifold,
\be
d^n x = \Big(\frac{1+x^2}{2}\Big)^n d\Omega 
\label{conformal}
\ee
naturally emerges in (\ref{k1}) and (\ref{k2}). Relation (\ref{k1}) hows that the product of the Killing vectors with repeated `a' indices yields, up to the conformal factor, the induced metric. It is exactly the induced metric only in 
$n=2$, which is a special case. The other relation can be interpreted as the transversality condition emanating from (\ref{s8}).
For computing derivatives involving Killing vectors, a particularly useful identity is given by,
\be
K_a^\mu\p_\mu K_a^\nu = (2-n)\Big({{1+x^2}\over 2}\Big) x_\nu
\label{k3}
\ee

The relation (\ref{s18b}) shows that the $(n+1)$-th component 
is just given by the dilatation (scaling), while the other 
components (given by (\ref{s18a}) involve the special conformal transformations and the translations. 

\bigskip

\section{Covariant Formulation of  Vector Gauge Theory}

\bigskip

In this section we discuss the manifestly $O(n+1)$-covariant formulation of a nonabelian vector gauge theory.
The theory is obtained by stereographically projecting the usual theory defined on the $n$-dimensional Euclidean plane 
on to the unit hypersphere embedded in $(n+1)$-dimensional Euclidean space. 
Results are known for the specific cases
of two and four dimensions \cite{A, DS, JR, NS}. 
We present the analysis for any $n$-dimensions. This is the first use of the 
generalisation effected by working in terms of the Killing vectors. 

   The pure Yang-Mills theory on the Euclidean space is governed by the standard Lagrangian,
\be
{\cal L}= -{1\over 4}F_{\mu\nu}^i F_{\mu\nu}^i
\label{v1}
\ee
where the field tensor is given by,
\be
F_{\mu\nu}^i=\p_\mu A_\nu^i - \p_\nu A_\mu^i +f^{ijk} A_\mu^j A_\nu^k
\label{v2}
\ee
and $f^{ijk}$ are the structure constants of the gauge group.

To define this theory on the hypersphere, the map among the potentials is first stated,
 \be
 \hat A_a^i = K_a^\m A_\m^i + r_a\phi^i
 \label{v3}
 \ee
which is a generalisation of (\ref{s7}). The scalar field (with $i$-multiplets) is included for the sake
of completeness, but will be discarded subsequently. These fields are the normal components of 
the potentials,
\be
\phi^i = r_a \hat A_a^i
\label{v4}
\ee
and are the analogues of (\ref{s9}).
They have a role in defining the shift transformations on the hypersphere. In order to see this, note
that the usual gauge tranformations are given by,
\be
\delta A_\mu^i =\p_\mu \lambda^i + f^{ijk} A_\mu^j \lambda^k
\label{v5}
\ee
The gauge transformations of the fields $(\hat A_a^i)$ on the hypersphere
are thus derivable from (\ref{v3}),
\be
\delta \hat A_a^i =K_a^\mu\p_\mu \lambda^i + f^{ijk}( \hat A_a^j - r_a r_b A_b^j )\lambda^k
+r_a \Lambda^i 
\label{v6}
\ee
where the definition (\ref{v4}) for the scalar fields has been used and we have set $ \delta\phi^i=\Lambda^i$.
It can be put in a more tractable form by introducing the angular
momentum operator,
\be
L_{ab}=r_a p_b-r_b p_a= -i(r_a\p_b -r_b\p_a)
\label{v7}
\ee
By using (\ref{s0}) and (\ref{s1}) it is possible to express the derivatives in terms of
those occurring in the plane and it is found that,
\be
L_{ab}= -i(r_a K_b^\mu -r_b K_a^\mu)\p_\mu
\label{v8}
\ee
Contracting by $r_a$ and using the tranversality condition (\ref{s8})
yields,
\be
r_a L_{ab} = -i K_b^\mu\p_\mu
\label{v9}
\ee
The transformation law (\ref{v6}) is thus expressed as,
\be
\delta \hat A_a^i = ir_b L_{ba}\lambda^i + f^{ijk}( \hat A_a^j - r_a r_b A_b^j )\lambda^k
+r_a \Lambda^i 
\label{v10}
\ee
The transformation of the scalars is completely arbitrary since the defining relation
(\ref{v4}) does not constrain $\Lambda^i$ in any way, as may be easily seen by taking
the variations on both sides of that equation. In fact it corresponds to a shift symmetry,
while the gauge symmetry is governed by the parameter $\lambda^i$. To compare with the 
transformation law given in the literature, it is necessary to consider the potentials
defined on the tangent plane of the hypersphere which, as has already been discussed, 
corresponds to setting the scalar field (\ref{v4}) to zero. Then the gauge transformation
is given by the first two terms in (\ref{v10}), which agrees with the known form \cite{JR}.
We have however provided a systematic derivation of this gauge transformation.

Although the infinitesimal gauge transformations were considered, it is quite simple to 
construct the finite version. If the ordinary potential transforms as,
\be
A_\mu' = U^{-1} (A_\mu +\p_\mu )U
\label{gt1}
\ee
then the projected potential transforms as,
\be
\hat A_a' = K_a^\m A_\mu'=  U^{-1} (\hat A_a +i r_b L_{ba} )U
\label{gt2}
\ee
obtained by using (\ref{v3}) and (\ref{v9}).

The next issue concerns a suitable mapping of the field tensors so that the 
appropriate actions may be obtained. Also, from now on we set the scalar field
to zero to facilitate comparison with existing results. Since translations on a plane are 
equivalent to rotations on a sphere, usual derivatives should be replaced by angular 
derivatives. The field tensor on the sphere is therefore a three index object.
It should thus be mapped to the field tensor on the flat space by the following
operation involving the Killing vectors,
\be
\hat F_{abc}^i= \Big(r_a K_b^\mu K_c^\nu +r_b K_c^\mu K_a^\nu + r_c K_a^\mu K_b^\nu\Big)F_{\mu\nu}^i
\label{v11}
\ee
so that symmetry properties under exchange of the indices is correctly preserved.

To show that this definition is equivalent to that followed in the literature \cite{JR},
\be
\hat F_{abc}^i = \Big(i L_{ab}\hat A_c^i + f^{ijk}r_a \hat A_b^j \hat A_c^k\Big) + c.p.
\label{v12}
\ee
(where $c.p.$ stands for the other pair of terms involving cyclic permutations in $a, b, c$), equations 
(\ref{v3}) and (\ref{v8}) are used to simplify (\ref{v12}),
\be
\hat F_{abc}^i = \Big(r_a K_b^\mu-r_b K_a^\mu\Big)\p_\mu\Big(K_c^\nu A_\nu^i\Big)
+f^{ijk}r_a\Big(K_b^\nu A_\nu^j\Big)
\Big(K_c^\mu A_\mu^k\Big) + c.p.
\label{v13}
\ee
The derivatives acting on the Killing vectors sum up to zero on account of the identity,
\be
\Big(r_a K_b^\mu-r_b K_a^\mu\Big)\p_\mu K_c^\nu  +c.p. = 0
\label{v14}
\ee
The derivatives acting on the potentials, together with the other pieces, combine to
reproduce (\ref{v11}), thereby completing the proof.

The components of the field tensor are easily read-off from (\ref{v11}),
\be
\hat F_{\mu\nu\sigma}^i= {{1+x^2}\over 2}\Big(x_\mu F_{\nu\sigma}^i 
+ x_\nu F_{\sigma\mu}^i + x_\sigma F_{\mu\nu}^i \Big)
\label{v15}
\ee
and,
\be
\hat F_{\mu\nu (n+1)}^i= 
{{1+x^2}\over 2}\Big( {{1-x^2}\over 2} F_{\mu\nu}^i - x_\mu x_\rho F_{\nu\rho}^i - x_\nu 
x_\rho  F_{\rho\mu}^i \Big)
\label{v16} 
 \ee
which is valid for any dimensions. For the particular examples in two and four
dimensions only, these were given in \cite{e}. Indeed 
the two dimensional case is special where only the
second component survives, which simplifies to,
\be
\hat F_{123}^i = ({{1+x^2}\over 2})^2 F_{12}^i
\label{v17}
\ee
Having obtained the map of the field tensor, it is straightforward to obtain the
action. Taking the repeated product of the field tensor (\ref{v11}) and using the
transversality of the Killing vectors, we get,
\be
\hat F_{abc}^i\hat F_{abc}^i= 3\Big(K_a^\mu K_a^\nu K_b^\lambda K_b^\rho\Big)F_{\mu\nu}^i F_{\lambda\rho}^i
\label{v18}
\ee
Finally, using (\ref{k1}), we obtain,
\be
\hat F_{abc}^i\hat F_{abc}^i= 3 \Big({{1+x^2}\over 2}\Big)^4 F_{\mu\nu}^i F_{\mu\nu}^i
\label{v19}
\ee
a result that is valid in any dimensions. In particular, on $S^4$, the conformal factor exactly cancels
and the actions on the flat space and the hypersphere are identified as,
\be
S =  -{1\over 4}\int d^4x F_{\mu\nu}^i F_{\mu\nu}^i =  -{1\over 12}\int d\Omega\hat F_{abc}^i\hat F_{abc}^i
\label{v20}
\ee
The lagrangian following from this action,
\be
{\cal L}_\Omega =-{1\over 12}\hat F_{abc}^i\hat F_{abc}^i
 \label{v21}
\ee
 is taken as the starting point of all 
computations on the hypersphere.

The inverse relations are obtained on using the properties of the Killing vectors. 
For example, multiplying (\ref{v3}) by $K_a^\nu$ and using (\ref{k1}), it follows,
\be
A_\nu^i = \Big({2\over {1+x^2}}\Big)^2 K_a^\nu\hat A_a^i
\label{v22}
\ee
It is possible to show, after some amount of algebra, that this relation is equivalent
to (\ref{s5}).
Likewise, contracting (\ref{v11}) by $r_a$ and using the transversality of the Killing vectors, we obtain,
\be
r_a\hat F_{abc}^i= \Big(K_b^\mu K_c^\nu \Big)F_{\mu\nu}^i
\label{v23}
\ee
from which, proceeding as before in deriving (\ref{v22}), we obtain the final form,
\be
F_{\mu\nu}^i = \Big({2\over {1+x^2}}\Big)^4 r_a K_b^\mu K_c^\nu \hat F_{abc}^i
\label{v24}
\ee

\bigskip

\subsection{Matter Fields and Killing Spinors}

\bigskip

The inclusion of the matter sector is also done with the help of the Killing vectors. 
Form invariance of the interaction requires that,
\be
\int dx (j_\mu^i A_\mu^i) = \int d\Omega (\hat j_a^i \hat A_a^i)
\label{m1}
\ee
where $j_\mu$ and $\hat j_a$ are the currents in the two decsriptions.
It is clear therefore that the currents are also mapped by a relation similar to
(\ref{v3}). However since the measure is given by (\ref{conformal}), the currents will
involve the conformal factor, depending on the dimensions. For the case of two dimensions,
it is exactly identical to (\ref{v3}),
\be
 \hat j_a^i = K_a^\m j_\m^i 
 \label{m2}
 \ee
 while for four dimensions, the conformal factor appears,
\be
 \hat j_a^i =\Big({{1+x^2}\over 2}\Big)^2 K_a^\m j_\m^i 
 \label{m3}
 \ee
Higher powers of the conformal factor occur for higher dimensions. Likewise,
the axial vector current is also defined. These forms
of the fermionic current have been postulated in the literature \cite{A}, particularly in
the discussion of anomalies and their connection with  index theorems \cite{NS}.
In the language of the Killing vectors the anomaly equation
can be expressed in a very economical manner,
as analysed below.

An essential ingredient in the discussion on anomalies is the completely antisymmetric
tensor $\e_{\mu\nu\lambda\rho....}$ whose value is the same in all systems. It is necessary
to define the analogue of this tensor  $\e_{abcd...}$ on the hypersphere.
We adopt the same rule (\ref{v11}) used for defining the antisymmetric field tensor.
However there is a slight subtlety. Strictly speaking, this Levi-Civita epsilon is a tensor density.
Hence its transformation law is modified by appropriate weight factors. For two dimensions, 
it is given by,
\be
\e_{abc}=\Big({2\over {1+x^2}}\Big)^2  
\Big(r_a K_b^\mu K_c^\nu +r_b K_c^\mu K_a^\nu + r_c K_a^\mu K_b^\nu\Big)\e_{\mu\nu}
\label{m4}
\ee
while in four dimensions it is given by,
\be
\e_{abcde}=\Big({2\over {1+x^2}}\Big)^4  
\Big(r_a K_b^\mu K_c^\nu K_d^\lambda K_e^\rho + cyclic\,\,\,\, permutations\,\,\,in\,\,\,(a, b, c, d, e) \Big)\e_{\mu\nu\lambda\rho}
\label{m5}
\ee
It is possible to verify the above relation by an explicit calculation, taking the convention
that both the epsilons are $+1 (-1)$ for any even (odd) permutation of distinct entries $(1, 2, 3,... )$ in that order.
Similar extension to other higher dimensions is straightforward.

The axial $U(1)$-anomaly in two and four dimensions is known to be given by,
\be
\p_\mu j_{\mu 5} = {1\over 2\pi}\e_{\mu\nu}F_{\mu\nu}
\label{anomaly2}
\ee
\be
\p_\mu j_{\mu 5} = {1\over {16\pi^2}}\e_{\mu\nu\lambda\rho}F_{\mu\nu}F_{\lambda\rho}
\label{anomaly4}
\ee

Using (\ref{v9}) and the definition of the currents (\ref{m2}), (\ref{m3}), it is possible to obtain the
identification,
\be
i r_a L_{ab} \hat j_{b5} = \Big({{1+x^2}\over 2}\Big)^2 \p_\mu j_{\mu 5}
\label{m8}
\ee 
for two dimensions and,
\be
i r_a L_{ab} \hat j_{b5} = \Big({{1+x^2}\over 2}\Big)^4 \p_\mu j_{\mu 5}
\label{m9}
\ee 
for four dimensions. In getting at the final result, use was made of the identity
(\ref{k3}). It is easy to extend this identification for higher dimensions, with
appropraite increase in the powers of the weight factor.

The explicit expressions for the anomaly are also identified with the minimum of effort.
For two dimensions it trivially follows from (\ref{v17}),
\be
\e_{abc}\hat F_{abc}= 6\hat F_{123}= 3\Big({{1+x^2}\over 2}\Big)^2\e_{\mu\nu}F_{\mu\nu}
\label{m10}
\ee
For four dimensions, using (\ref{v11}) and (\ref{m5}), it follows that,
\be
r_a\e_{bcdef}\hat F_{abc}\hat F_{def}= 
3\Big({{1+x^2}\over 2}\Big)^4\e_{\mu\nu\lambda\rho}F_{\mu\nu}F_{\lambda\rho}
\label{m11}
\ee
The structural similarity among
the results in two and four dimensions is  obvious. 
Once again, extension to other dimensions is equally
clear and can be done naturally.
Inclusion of the nonabelian gauge group also poses no problems and is done by 
inserting suitable indices. The weight factors cancel out from both sides of the 
anomaly equation, which now takes its familiar form \cite{NS, DW},
\be
i r_a L_{ab} \hat j_{b5} = {1\over {48\pi^2}}r_a\e_{bcdef}\hat F_{abc}\hat F_{def}
\label{m12}
\ee

Before concluding this section, we derive the map among the basic fermion fields.  
From the transversality of the Killing
vectors it is clear that the  currents, as defined in  (\ref{m2}) or (\ref{m3}) 
(including  other dimensions), must satisfy,
\be
r_a \hat j_a =0
\label{m13}
\ee
Thus, modulo a normalisation, the fermionic current on the hypersphere must be given by,
\be
\hat j_a =\hat \psi^\dagger (\delta_{ab}- r_a r_b)\sigma_b \hat \psi
\label{m14}
\ee
where $\sigma_b$ are a set of hermitian matrices generating the Clifford algebra,
$\{\sigma_a, \sigma_b\}= 2\delta_{ab}$. For $S^2$ embedded in $R^3$, these are the standard Pauli matrices.
Also note that since the map between the fermion variables will be linear, the normalisation
may be absorbed and thus set to unity.

Let us now perform the analysis for $S^2$. The examples of $S^4$ or other hyperspheres may be done
analogously. Using (\ref{m2}) and the identity (\ref{k2}), we obtain,
\be
\Big({2\over {1+x^2}}\Big)^2 \hat \psi^\dagger K_{a}^\lambda K_b^\lambda \sigma_b \hat \psi=
\psi^\dagger K_a^\mu\gamma_\mu\psi
\label{m15}
\ee
where the Euclidean gamma matrices are given by $\gamma_\mu =\e_{\alpha\mu}\sigma_\alpha.$
This may be simplified as,
\be
 \hat \psi^\dagger K_{a}^\lambda  \sigma_a \hat \psi= \Big({{1+x^2}\over 2}\Big)^2
\psi^\dagger \gamma_\lambda \psi
\label{m16}
\ee
Now defining the map between the fermi fields by,
\be
\psi = \Big({2 \over {1+x^2}}\Big) W\hat \psi
\label{m17}
\ee
where $W$ is the transformation matrix, we get the following condition,
\be
 K_{a}^\lambda  \sigma_a= W^\dagger \gamma_\lambda W
\label{m18}
\ee
To solve this equation it becomes convenient to square both sides,
\be
2\Big(\frac{1 +x^2}{2}\Big)^2 = ( W^\dagger \gamma_\lambda W)
( W^\dagger \gamma_\lambda W)
\label{square}
\ee
obtained on using the identity (\ref{k2}) and properties of the Pauli matrices.
Since $\gamma_\lambda\gamma_\lambda =2$, it leads to the following simplification,
\be
W W^\dagger = W^\dagger W = \Big(\frac{1 +x^2}{2}\Big)
\label{square1}
\ee
This suggests the following solution,
\be
W={1\over {\sqrt 2}}(1-i\gamma_\mu x_\mu)
\label{m19}
\ee
which is checked from (\ref{m18}) by using 
the known forms for the Killing vectors.

Putting this in (\ref{m17}), the result for the mapping of the fermion
fields quoted in the literature \cite{A, JR, NS} is reproduced. For four
dimensions, expectedly the only change comes in the weight factor, which now occurs
with a power of 2,
\be
\psi = \Big({2 \over {1+x^2}}\Big)^2 W\hat \psi
\label{m17a}
\ee
 Similar features hold in other higher dimensions. 
The matrix $W$, on the other hand, is fundamental, as is now elucidated.

Equation (\ref{m18}) shows that the conformal Killing vectors are expressed as a bilinear 
combination of $W's$. Consequently these are related to the conformal Killing spinors.
As is known \cite{LPR}, the Killing spinors on $S^n$ are expressed as the product of a 
matrix with an arbitrary  constant spinor. In the case of  $S^2$, for example,
 the Killing spinor  is given by 
$\epsilon = \Omega \left(\begin{array}{c}a\\
b\end{array}\right)
$. A bilinear combination yields the Killing vector $K^\mu = \epsilon'^\dagger
\gamma^\mu \epsilon$, where $\epsilon' = \Omega \left(\begin{array}{c}a'\\
b'\end{array}\right)
$. The transformation matrix $\Omega$ multiplying the constant spinor is
found by solving a consistency condition.  The formalism here naturally 
yields the conformal Killing spinors with $W$ replacing the matrix $\Omega$.
Usually these matrices are given in terms of the Euler angles, but that can be
trivially obtained here by using the formulas related to stereographic projection
and finally passing over to the polar variables. In particular, using the parametrisation
(\ref{w4}) used later, the structure for the matrix turns out to be,
\be
\sqrt 2 W=
\left(\begin{array}{clcr}
1 & e^{-i\phi}tan\frac{\theta}{2}\\
-e^{i\phi}tan\frac{\theta}{2} & 1
\end{array}\right)
\label{matrix}
\ee

For Killing spinors on $S^n$, the matrix $\Omega$ turns out to be unitary{\footnote{Although this
was not discussed in \cite{LPR}, it can be proved from their results}}. Hence, using (\ref{square1}),
it transpires that the matrices $W$ and $\Omega$ are, up to a conformal factor, unitarily 
equivalent.

Relation (\ref{m17}) shows that the mapping among the matter fields is effected by
the conformal Killing spinors. This complements the mapping in the gauge sector done by the
conformal Killing vectors.

\bigskip

\section{Covariant Formulation of Antisymmetric Tensor Gauge Theory}

\bigskip

The general formalism developed so far is particularly suited for obtaining a 
covariant formulation of $p$-form gauge theories. Here we discuss it for the 
second rank antisymmetric tensor gauge theory. Also, there are some features
which distinguish it from the analysis for the  vector gauge theory.
The extension for higher forms 
is obvious. Both abelian and nonabelian theories will be considered. To set
up the formulation it is convenient to begin with the abelian case which can
be subsequently generalised to the nonabelian version. The action for a free
2-form gauge theory in flat $n$-dimensional  Euclidean space is given by \cite{KR},
\be
S= -{1\over {12}}\int d^n x F_{\mu\nu\rho}F_{\mu\nu\rho}
\label{t1}
\ee
where the field strength is defined in terms of the basic field as,
\be
 F_{\mu\nu\rho}= \p_\mu B_{\nu\rho}+ \p_\nu B_{\rho\mu}+ \p_\rho B_{\mu\nu}
\label{t2}
\ee
The infinitesimal gauge symmetry is given by the transformation,
\be
\d B_{\mu\nu} = \p_\mu \Lambda_\nu -\p_\nu \Lambda_\mu
\label{t3}
\ee
which is reducible since it trivialises for the choice $\Lambda_\mu = \p_\mu\lambda$.

It is sometimes useful to express the action (or the lagrangian) in a first order form
by introducing an extra field,
\be
{\cal L}= -{1\over {8}}\e_{\mu\nu\rho\sigma} F_{\mu\nu}B_{\rho\sigma}+ {1\over 8}A_\mu A_\mu
\label{t4}
\ee
where the $B\wedge F$ term involves the field tensor,
\be
F_{\mu\nu}=\p_\mu A_\nu - \p_\nu A_\mu
\label{t5}
\ee
Eliminating the auxiliary $A_\mu$ field by using its equation of motion, the previous
form (\ref{t1}) is reproduced. The gauge symmetry is given by (\ref{t3}) together with
$\d A_\mu=0$. The first order form is ideal for analysing the nonabelian theory.

To express the theory on the hypersphere, the mapping of the tensor field is first given.
From the previous analysis, it is simply given by,
\be
\hat B_{ab} = K_a^\mu K_b^\nu B_{\mu\nu} +r_a\phi_b - r_b\phi_a
\label{t6}
\ee
The tensor field with the latin indices is defined on the hypersphere while those with
the greek symbols are the usual one on the flat space. There are additional vector fields
$\phi_a$, which are the analogues of the scalar field $\phi$, given in (\ref{v3}).
However, there is a point of distinction 
from the vector theory, which is explained by counting the degrees of freedom. The
vector field on $S^n$ embedded in $R^{n+1}$ had $(n+1)$-components that were expressed in
terms of the $n$-components of the usual vector on $R^n$ plus one extra scalar degree of
freedom. Subsequently by constraining the vector field to lie on the tangent plane of the
hypersphere, the scalar field could be set to zero. For a rank 2 tensor on the hypersphere
there are ${{(n+1)n}\over 2}$ components, while on the flat space it has ${{n(n-1)}\over 2}$
components. Their difference is $n$. Since $\phi_a$ has $(n+1)$ components, all of them cannot
be independent. We have to impose an additional constraint; their normal component is set to
zero,
\be
r_a\phi_a=0
\label{t7}
\ee
so that there is a correct matching of the degrees of freedom.
By contracting (\ref{t6}) with $r_a$, using the above equation and the transversality of the
Killing vectors, we get,
\be
r_a \hat B_{ab}= \phi_b
\label{t8}
\ee
showing that $\phi_a$ is the normal component of the tensor field, expectedly
satisfying (\ref{t7}). In analogue with the 
vector theory it is possible to do away with this field completely by requiring that 
the tensor field resides on the tangent plane of the hypersphere, in which case it is given by,
\be
\hat B_{ab} = K_a^\mu K_b^\nu B_{\mu\nu} 
\label{t9}
\ee
This is written in component notation by using the explicit form  for the Killing vectors 
given in (\ref{s18a}) and (\ref{s18b}),
\be
\hat B_{\m\n} =\frac{1+x^2}{2}\Big(\frac{1+x^2}{2} B_{\m\n} - x_\rho x_\n B_{\m\rho}
-x_\rho x_\m B_{\rho\n}\Big)
\label{t10}
\ee
and,
\be
\hat B_{\m(n+1)} = -\frac{1+x^2}{2} x_\rho B_{\m\rho}
\label{t11}
\ee
These are the analogues of (\ref{s2}) and (\ref{s3}). The inverse relation is given by,
\be
\Big(\frac{1+x^2}{2}\Big)^4 B_{\m\n} = K_a^\m K_b^\n \hat B_{ab}
\label{t12}
\ee
which may also be put in the form,
\be
\Big(\frac{1+x^2}{2}\Big)^2 B_{\m\n} =  \hat B_{\m\n} +x_\m \hat B_{\n(n+1)}- x_\n \hat B_{\m(n+1)}
\label{t13}
\ee
which is the direct analogue of (\ref{s5}).

Next, the gauge transformations are discussed. From (\ref{t3}), the defining relation (\ref{t9}) and the 
angular momentum operator (\ref{v9}), infinitesimal transformations are given by
{\footnote{If the additional $\phi_a$-fields had been retained, the shift symmetry
$\d_s \hat B_{ab}= r_a\d\phi_b - r_b \d\phi_a$ would have been found,
exactly as happened in the previous example.}},
\be
\d \hat B_{ab}= i r_c\Big(K_b^\m L_{ca} -K_a^\m L_{cb}\Big)\Lambda_\m
\label{t14}
\ee
In this form the expression is not manifestly covariant. This may be contrasted with
(\ref{v10}) which has this desirable feature. The point is that an appropriate map
of the gauge parameter is necessary. In the previous example the gauge parameter was a scalar
which retained its form. Here, since it is a vector, the required map is provided by a relation
like (\ref{v3}), so that,
\be
\hat\Lambda_a = K_a^\mu \Lambda_\mu
\label{t14.1}
\ee
Pushing the Killing vectors through the angular momentum operator and using the above 
map yields,
\be
\d \hat B_{ab}= i r_c\Big( L_{ca}\hat\Lambda_b - L_{cb}\hat\Lambda_a \Big)
+i r_c \Lambda_\mu
\Big(L_{cb} K_a^\mu - L_{ca} K_b^\mu \Big)
\label{t14.2}
\ee
The last bracket can be simplified leading to the cherished expression,
\be
\d \hat B_{ab}= i r_c\Big( L_{ca}\hat\Lambda_b - L_{cb}\hat\Lambda_a \Big)
- r_a\hat\Lambda_b + r_b\hat\Lambda_a
\label{t14.3}
\ee
This transformation could have been deduced from other arguments too. On grounds of covariance and
symmetry, the expression within the brackets is expected. The remainder is necessary to satisfy
the tranversality condition $\d (r_a\hat B_{ab})=0$, which is inherent in the construction.
This condition follows from (\ref{t14.3}) on use of the identity,
\be
\hat\Lambda_a +ir_b r_c L_{ba}\hat\Lambda_c = 0
\label{t14.4}
\ee
obtained on repeated use of $r_a\hat\Lambda_a =0$ and relations derived by successively
differentiating it with operators like $\p_a, \p_b etc$.

It is also reassuring to note that (\ref{t14.3}) manifests the reducibility of the
gauge transformations. Since $\Lambda_\mu=\p_\mu\lambda$ leads to a trivial gauge
transformation in 
flat space, it follows from (\ref{t14.1}) that the corresponding feature should be present
in the hyperspherical formulation when,
\be
\hat\Lambda_a = i r_c L_{ca}\lambda
\label{t14.5}
\ee
It is easy to check that with this choice, the gauge transformation (\ref{t14.3}) 
trivialises; i.e. $\d \hat B_{ab}=0$.

The field tensor on the hypersphere is constructed from the usual one given in (\ref{t2}).
Since the Killing vectors play the role of the metric in connecting the two surfaces, this
expression is given by a natural extension of (\ref{v11}),
\be
\hat F_{abcd}= \Big(r_a K_b^\mu K_c^\nu K_d^\rho+r_b K_c^\mu K_a^\nu K_d^\rho + r_c K_d^\mu K_a^\nu
K_b^\rho +r_d K_a^\mu K_c^\nu K_b^\rho\Big)F_{\mu\nu\rho}
\label{t15}
\ee
Note that cyclic permutations have to taken carefully since there is an even number of
indices.
In terms of the basic variables, the field tensor is expressed as,
\be
\hat F_{abcd}= i\Big(L_{ab}\hat B_{cd} +L_{bc}\hat B_{ad} + L_{bd}\hat B_{ca} + L_{ca}\hat B_{bd} +
L_{da}\hat B_{cb}+
L_{cd}\hat B_{ab}\Big)
\label{t16}
\ee

To show that (\ref{t15}) is 
equivalent to (\ref{t16}), the same strategy as before, is 
adopted. Using the definition of the angular 
momentum (\ref{v8}), (\ref{t16}) is simplified
as,
\be
\hat F_{abcd}=\Big(r_a K_b^\mu - r_b K_a^\mu\Big)\p_\mu\Big( K_c^\nu K_d^\sigma B_{\nu\sigma} \Big)
+............
\label{t17}
\ee
where the carets denote the inclusion of other similar 
(cyclically permuted) terms. Now there are two types
of contributions. Those where the derivatives act on the Killing vectors and those where
they act on the fields. The first class of terms cancel out as a consequence of an identity
that is an extension of (\ref{v14}). The other class combines to reproduce (\ref{t15}).

The action on the hypersphere is now obtained by first taking a repeated product of the field
tensor (\ref{t15}). Using the properties of the Killing vectors, this yields,
\be
\hat F_{abcd}\hat F_{abcd}=4\Big(\frac {1+x^2}{2}\Big)^6 F_{\mu\nu\rho}F_{\mu\nu\rho}
\label{t18}
\ee
From the definition of the flat space action (\ref{t1}) and the volume element (\ref{conformal}),
it follows that the above identification leads to the hyperspherical action (for $n=4$),
\be
S_\Omega =-\frac{1}{48}\int d\Omega \Big(\frac{2}{1+x^2}\Big)^2\hat F_{abcd}\hat F_{abcd}
\label{t19}
\ee

Thus, up to a conformal factor, the corresponding lagrangian is given by,
\be
{\cal L}_\Omega =-\frac{1}{48}\hat F_{abcd}\hat F_{abcd}
\label{t20}
\ee
Only in six dimensions, the conformal factor does not occur. This would be the analogue of
the vector theory in four dimensions.

By its very construction this lagrangian would be invariant under the gauge transformation
(\ref{t14.3}). There is however another type of gauge symmetry 
which does not seem to have any analogue in the flat space. 
To envisage such a  possibility, consider a  transformation of the
type {\footnote{Recently such a transformation was considered in \cite{M},
though only on $S^4$}},
\be
\delta \hat B_{ab}=L_{ab}\lambda
\label{t21}
\ee
which  could be a meaningful gauge symmetry
operation on the hypersphere. However, in flat space, it
leads to a trivial gauge transformation. To see this explicitly, consider the effect of
(\ref{t21}) on (\ref{t12}),
\be
\Big(\frac{1+x^2}{2}\Big)^4 \delta B_{\m\n} = K_a^\m K_b^\n L_{ab}\lambda
\label{t22}
\ee
Inserting the expression for the angular momentum from (\ref{v7}) and
exploiting the transversality (\ref{s8}) of the Killing vectors, it follows that,
\be
 \delta B_{\m\n} = 0
\label{t23}
\ee
thereby proving  the statement. To reveal that (\ref{t21})
indeed leaves the lagrangian (\ref{t20}) invariant, it is desirable 
to recast it in the form,
\be
{\cal L}_\Omega =\frac{1}{32}\hat \Sigma_{a}\hat \Sigma_{a}
\label{t24}
\ee
where,
\be
\hat \Sigma_a = \e_{abcde}L_{bc}\hat B_{de}
\label{t25}
\ee
Under the gauge transformation (\ref{t21}), a simple algebra shows that $\delta \hat \Sigma_a = 0 $ and hence 
the lagrangian remanins invariant.

The inclusion of a nonabelian gauge group is feasible. Results follow logically
from the abelian theory with suitable insertion of the nonabelian indices. As
remarked earlier it is useful to consider the first order form (\ref{t4}). The
lagrangian is given by its generalisation \cite{FT},
\be
{\cal L}= -{1\over {8}}\e_{\mu\nu\rho\sigma} F_{\mu\nu}^i B_{\rho\sigma}^i + {1\over 8}A_\mu^i A_\mu^i
\label{t26}
\ee
where the nonabelian field strength has already been defined in (\ref{v2}). It is
gauge invariant under the nonabelian generalisation of (\ref{t3}) with the 
ordinary derivatives replaced by the covariant derivatives with respect to the potential 
$A_\mu$, and $\d A_\mu^i = 0$. By the help of our equations it is straightforward to
project this lagrangian on the hypersphere. For instance, the corresponding gauge 
transformations look like,
\be
\d \hat B_{ab}^i = i r_c\Big( L_{ca}\hat\Lambda_b^i - L_{cb}\hat\Lambda_a^i \Big)
- r_a\hat\Lambda_b^i + r_b\hat\Lambda_a^i +f^{ijk}(\hat A_a^j \hat\Lambda_b^k
- \hat A_b^j \hat\Lambda_a^k)
\label{t27}
\ee
and so on.

Matter fields may be likewise defined. The fermion current $j_{\mu\nu}$ will
be defined just as the two form field, except that conformal weight factors appear,
so that form invariance of the interaction is preserved,
\be
\int dx (j_{\mu\nu}^i B_{\mu\nu}^i) = \int d\Omega (\hat j_{ab}^i \hat B_{ab}^i)
\label{t28}
\ee
quite akin to (\ref{m1}).

\bigskip

\section{Killing Vectors and Group Contraction}

Now we consider the case of $S^2$ in some details. Since $SO(3)$ as a group of
transformation is the symmetry group of a surface of a sphere,  the Killing vectors discussed
here must be  equivalent to the angular momentum operators, which are the generators of this group.
This will be shown explicitly. Furthermore, by using the technique of group contraction,
two of these Killing vectors reduce to the translation generators on the plane, while the third
remains invariant. Together these constitute the generators of $E_2$, which is the symmetry
group of the plane. Thus the familiar contraction of $SO(3)$ to $E_2$ is revealed.

Since we have specialised to $S^2$ it is convenient to recast the three components of
the angular momentum (\ref{v8}) as,
\be
L_a = \frac{1}{2}\e_{abc} L_{bc}
\label{w1}
\ee
By using an identity,
\be
\e_{abc}r_b K_c^\mu = \e_{\mu\nu}K_a^\nu
\label{w2}
\ee
it is possible to write (\ref{w1}) directly in terms of the rotated Killing vectors,
\be
L_a =-i \e_{\mu\nu}K_a^\nu\p_\mu
\label{w3}
\ee

The spherical coordinates are expressed in terms of the polar angles,
\be
r_1=\sin\theta \cos\phi ,\,\,\, r_2=\sin\theta \sin\phi ,\,\,\,r_3=\cos\theta 
\label{w4}
\ee
Using this transformation, the explicit forms for the Killing vectors (\ref{s18a}), (\ref{s18b}),
and the relation (\ref{inverse}), the following result is derived from (\ref{w3}),
\begin{eqnarray}
L_1 &=& i \sin\phi\,\,\p_\theta +i \cot\theta \,\,\cos\phi \,\,\p_\phi\\
L_2 &=& -i \cos\phi\,\,\p_\theta +i \cot\theta\,\, \sin\phi \,\,\p_\phi\\
L_3 &=& -i  \p_\phi
\label{map}
\end{eqnarray}
which are the familiar angular momentum operators generating the $SO(3)$ symmetry.

To discuss the contraction of $SO(3)$ to $E_2$, the $L_a$-operators in (\ref{w3}) are written
in terms of the planar coordinates,
\begin{eqnarray}
L_1& =& i\frac{x_1 x_2}{R}\p_1 +i\Big(\frac{R^2 +x^2}{2R}- \frac{x_1^2}{R}\Big)\p_2\\
L_2& =&- i\frac{x_1 x_2}{R}\p_2 -i\Big(\frac{R^2 +x^2}{2R}- \frac{x_2^2}{R}\Big)\p_1\\
L_3& =& i x_2\p_1 -i x_1\p_2
\label{contract}
\end{eqnarray}
where the radial coordinate has been explicitly inserted. Taking the large radius limit so that
the sphere approximates to the plane (around the 3-axis), these identifications are obtained,
\be
Lt_{R\rightarrow\infty}\frac{2L_1}{R} = i\p_2 =- P_2
\label{w5}
\ee
and,
\be
Lt_{R\rightarrow\infty}\frac{2L_2}{R} =- i\p_1 = P_1
\label{w6}
\ee
while $L_3$ remains invariant. Here $P_1$ and $P_2$ are the usual translation generators on the plane.
Along with $L_3$ it comprises the $E_2$ group. 
It is easy to verify that the Lie algebra is preserved in the large $R$-limit. Thus,
\be
[P_1, P_2]= \frac{4}{R^2}[L_2, L_1] = 0 
\ee
\be
[P_1, L_3]= \frac{2}{R}[L_2, L_3] = i\frac{2L_1}{R}=- i P_2
\ee
\be
[P_2, L_3]= -\frac{2}{R}[L_1, L_3] = i\frac{2L_2}{R}= i P_1
\label{lie}
\ee
thereby reproducing the algebra of $E(2)$; $[P_\alpha, P_\beta]
= 0, [P_\alpha, L_3]=-i\epsilon_{\alpha\beta}P_\beta$.
This completes the demonstration of the group contraction.

\bigskip

\section{Conclusions}

\bigskip

We have discussed a manifestly $O(n+1)$ covariant formulation of vector and tensor
gauge theories. It was done  by mapping  the usual forms of these theories, defined
on the Euclidean flat surface, onto a unit hypersphere by the method of stereographic
projection. A distinctive feature was to provide a first principle
analysis of this mapping to abstract the relevant conformal
Killing vectors by solving the Cartan-Killing equation. The importance of these
Killing vectors lay in the fact that tensor forms constructed by taking their 
products acted like a metric connecting the results between the flat space
and the hypersphere. This essentially new ingredient was crucial for generalisations
to include higher form gauge theories or arbitrary dimensions.
Furthermore, in this formalism the results of a vector gauge theory 
became illuminating and were compactly
reproduced. 

  As stated above, an extension of the analysis to include the two (and higher) forms was possible. The
two form tensor gauge theory was worked out in details. Once again the transition from the
flat surface to the hypersphere was completely encapsuled in the conformal Killing vectors.
A gauge symmetry was found in the hyperspherical formulation that did not
have any analogue in the flat case.

Matter fields were suitably inducted in the formalism. Just as the mapping in the gauge sector
was done through  conformal Killing vectors, that among the basic matter fields was 
realised through conformal Killing spinors. An equation was found, revealing these Killing
vectors as bilinear combinations of these Killing spinors. From a knowledge of the Killing
vectors, 
an explicit structure for the conformal 
Killing spinors was obtained. It may be mentioned that there exist results for the Killing spinors
on $S^n$ \cite{LPR}. 
However these have to be calculated case by case and their explicit forms are known only
in a few low dimensional examples (like, $n=2, 3, 4$). Here a compact formula 
(\ref{m19}) for the conformal
Killing spinors was given, valid in any dimensions. Also, it was shown that the transformation
matrices defining the conformal Killing spinors and usual Killing spinors were unitarily
equivalent, modulo the conformal factor.

Finally, the Killing vectors obtained here were shown to induce the contraction of the $SO(3)$ 
group to $E(2)$, in the limit where the radius of $S^2$ was taken very large.

Regarding future prospects we presume that, apart from its use in discussing the
quantisation of such theories, it may find applications in the recent
discussions on noncommutative gauge theories defined on a fuzzy sphere 
\cite{WW, APS, I}. A possible approach  
is to stereographically project the models on the noncommutative plane onto the
fuzzy sphere. On account of the noncommutativity, there is an ordering problem which
makes  it difficult to extract closed form
expressions and recourse is sometimes taken to coherent states \cite{APS, I}.  
Although some results
exist for the projection of the coordinates, that for the gauge or matter fields remains
unknown. Based on a systematic approach using Killing vectors (and spinors) it may be
feasible to construct such a mapping.

\bigskip

{\Large{\bf Acknowledgments}}
This work was supported by a grant from the Physics Department, 
SungKyunKwan University, Korea. I thank members of this department for their gracious 
hospitality. Also, I express my thanks to Satoshi Iso and Keiichi Nagao for discussions
during my stay at KEK, Japan, where this work was initiated.

\end{document}